\documentclass[useAMS]{mn2e}
\usepackage{times,psfig}

\def\MSUN{\rm M_{\odot}}

\def\MSUNYR{\rm M_{\odot}\,yr^{-1}}

\def\MDOT{\dot{M}}

\newbox\grsign \setbox\grsign=\hbox{$>$} \newdimen\grdimen \grdimen=\ht\grsign
\newbox\simlessbox \newbox\simgreatbox
\setbox\simgreatbox=\hbox{\raise.5ex\hbox{$>$}\llap
     {\lower.5ex\hbox{$\sim$}}}\ht1=\grdimen\dp1=0pt
\setbox\simlessbox=\hbox{\raise.5ex\hbox{$<$}\llap
     {\lower.5ex\hbox{$\sim$}}}\ht2=\grdimen\dp2=0pt

\begin{document}
\title[]
{The late time evolution of Gamma-Ray Bursts:
ending hyperaccretion and producing flares}

\author[]
{\parbox[]{6.in} {Daniel Proga$^1$ and Bing Zhang$^1$ \\
\footnotesize        
$^1$ Department of Physics, University of Nevada, Las Vegas,
NV 89154, USA, e-mail: dproga@physics.unlv.edu; bzhang@physics.unlv.edu}
\date{Accepted  .      Received ; in original form }}
\maketitle

\label{firstpage}

\begin{abstract}
We consider the properties of a hyperaccretion model for gamma-ray bursts
(GRBs) at late times when the mass supply rate is expected to decrease
with time. We point out that the region in the vicinity of the accretor
and the accretor itself can play an important role in determining
the rate of accretion, and its time behavior, and ultimately the energy output.
Motivated by numerical simulations and theoretical results,
we conjecture that  the energy release can be repeatedly
stopped and then restarted by the magnetic flux accumulated
around the accretor.
We propose that the episode or episodes when the accretion resumes
correspond to X-ray flares discovered recently in a number of GRBs.
\end{abstract}


\section{Introduction}
Gamma-Ray Bursts (GRB) are generally believed to be powered by hyperaccretion
onto a compact, stellar mass object. The total amount of the available fuel
is considered to be the key factor determining the burst duration.
Within merger scenarios for short-duration GRBs, a neutron star (NS)
is  accreted onto another NS
or onto a stellar mass black hole (BH; e.g, Paczy\'nski 1986, 1991;
Eichler et al. 1989; Narayan et al. 1992; Fryer et al. 1999).
Within the collapsar model for long-duration GRBs,
up to 20 $\MSUN$ of a stellar envelope collapses onto the star's core
which is a NS or a BH (e.g., Woosley 1993; Paczy\'nski 1998;
MacFadyen \& Woosley 1999; Popham, Woosley, \& Fryer 1999; Proga et al. 2003).
For short- and long-duration events, the accretion
rate, $\MDOT_a$ must be of order of 1~$\MSUN~{\rm s^{-1}}$, yielding a duration
of less than a few seconds for the former and a duration as long as tens
to hundreds of seconds for the latter. These duration estimates are made
under the assumption
that all the available fuel is accreted during the GRB activity
at a time-averaged constant rate.

Recent GRB observations obtained with {\it Swift} motivate us to review
the above assumption and some other aspects of  GRB models.
In particular, early X-ray afterglow lightcurves of
nearly half of the long-duration GRBs show X-ray flares
(Burrows et al. 2005; Romano et al. 2006; Falcone et al. 2006). 
X-ray flares are also
found to follow the short-duration GRB 050724 (Barthelmy et al. 2005)
whose host galaxy is early-type, which is consistent with the merger
origin. The flares generally rise and fall rapidly, with typical
rising and falling time scales much shorter than the epoch when the
flare occurs. This time behavior strongly supports the ``internal'' origin
of the flares (Burrows et al. 2005; Zhang et al. 2006; Fan \& Wei 2005), 
in contrast to
the ``external'' origin of the power-law decay afterglows. The
internal model not only offers a natural interpretation of the rapid
rise and decay behavior of the flares, but also demands a very small
energy budget (Zhang et al. 2006). Within this picture, the data
require a {\it restart} of  the GRB central engine (i.e., a restart
of accretion).

Fragmentations in the collapsing star (King et al. 2005) or in the
outer parts of the accretion disc (Perna et al. 2006) have been
suggested to be responsible for the observed episodic flaring behavior.
These two flare models appeal to one of the basic ingredients of an accretion
powered engine -- the mass accretion rate -- and conjecture that
the episodic energy output is driven by  changes in the mass supply and
subsequently accretion rate. In this picture, the inner
part of the accreting system {\em passively} responds
to changes in the accretion flow at larger radii.

Here, we point out that the region in the vicinity of the accretor
and the accretor itself can play an important role of determining
the rate and time behavior of the accretion and the energy output.
In particular, we conjecture that  the energy release can be repeatedly
stopped and then restarted, provided the mass supply rate decreases with time
even if the decrease is smooth. For both merger and collapsar
GRB models, a decrease of the mass supply
rate is expected, especially in the late phase of activity,
because the mass density decreases with increasing radius.
In our model, we appeal to the fact that, as mass is being accreted onto
a BH, the magnetic flux is accumulating in the vicinity of the BH.
Eventually, this magnetic flux must become dynamically important
and affect the inner accretion flow, unless the magnetic field
is very rapidly diffused. In the remaining part of the paper
we list and discuss theoretical arguments and results from
a variety of numerical magnetohydrodynamic (MHD) simulations
of accretion flows that support our model. We also provide
analytic estimates to show that our model can quantitatively account
for the observed features of the flares.

\section[]{Magnetic model for GRBs and their flares}

\subsection{Insights from numerical models}

Generally, our model for the flares is based on the results from the numerical 
simulation of an MHD collapsar model for GRBs carried out by Proga et al. 
(2003) and the results from a number of simulations of 
radiatively inefficient accretion flows (RIAFs) onto a BH 
(Proga \& Begelman 2003, PB03 hereafter;
Igumenshchev, Narayan, \&Abramowicz 2003, INA03 hereafter).
Proga et al. (2003) performed  time-dependent axisymmetric MHD simulations of
the collapsar model.  These MHD simulations included a realistic equation
of state, neutrino cooling, photodisintegration of helium,
and resistive heating. The progenitor was assumed to be
spherically symmetric but with spherical symmetry broken by the 
introduction of a small, latitude-dependent angular momentum 
and a weak split-monopole magnetic field.
The main conclusion from the simulations is that,
within the collapsar model, MHD effects alone are able to launch,
accelerate and sustain a strong polar outflow.
The MHD outflow provides favorable initial conditions for the subsequent
production of a baryon-poor fireball 
(e.g., Fuller, Pruet \& Abazajian 2000;
Beloborodov 2003; Vlahakis \& K$\ddot{\rm o}$nigl 2003;
M\'{e}sz\'{a}ros 2002), or a magnetically dominated ``cold
fireball'' (Lyutikov \& Blandford 2002), though the specific
toroidal magnetic field geometry Proga et al. derived differs from some
of these models (e.g., Vlahakis \& K$\ddot{\rm o}$nigl 2003;
Lyutikov \& Blandford 2002).
The latest Swift UV-Optical Telescope (UVOT) observations indicate that
the early reverse shock emission is generally suppressed (
Roming et al. 2005), which is consistent with the suggestion that at least 
some GRBs are Poynting-flux-dominated outflows (Zhang \& Kobayashi 2005).

To study the extended GRB activity, one would like to follow the collapse
of the entire star. However, such studies are beyond current computer
and model limits. Therefore, we explore instead the implications of 
the published simulations and consider the physics of the collapsing star 
to infer the properties and physical conditions in the vicinity of a BH
during the late phase of evolution, i.e., when a significant
fraction of the total available mass is accreted.

The long time evolution of axisymmetric MHD accretion flows 
was studied by PB03 who explored simulations very similar to those
performed by Proga et al. (2003) but with much simpler micro physics
(i.e., an adiabatic equation of state, no neutrino cooling
or photodisintegration of helium). Proga et al. (2003) found that
despite the more sophisticated micro physics of the MHD collapsar
simulations the flow cooling is dominated by advection not
neutrino cooling.
As a result, the early phase of the time evolution, 
and the dynamics of the innermost flow, are very similar in both
the  RIAF simulations and the collapsar simulations. 
In particular, after an
initial transient behavior, the flow settles into a complex convolution
of several distinct, time-dependent flow components
including an accretion torus, its corona and outflow, and
an inflow and outflow in the polar funnel (see the left panel in Fig. 1 for
a schematic picture of such a flow).
The accretion through the torus is facilitated by the magnetorotational
instability (MRI, e.g., Balbus \& Halwey 1991) which also
dominates the overall dynamics of the inner flow. 

In the remaining part of the paper, we will assume that
the late evolution of the MHD collapsar simulations
is similar to the late evolution of the RIAFs simulations.
This assumption is justifiable because the flows in the collapsar and RIAFs
simulations are similar during the early phase of the evolution
(i.e., their dymanics and cooling are dominated by MRI and advection, 
respectively)

The late evolution of RIAFs shows that the torus accretion can
be interrupted for a short time by a strong poloidal magnetic field in
the vicinity of a BH. This result is the main motivation for this paper,
as it shows that  the extended GRB activity may be a result
of an accretion flow modulated by
the ``magnetic-barrier'' and gravity. Because this barrier halts
the accretion flow intermittently (see Figs.~6 \& 8 in PB03),
the accretion rate is episodic (see Fig.3 of PB03).
This potentially gives a natural mechanism
for flaring variability in the magnetic-origin models of GRBs as
we first mentioned in Fan, Zhang \& Proga (2005;
see the middle panel of Fig. 1 here, for a cartoon picture of
the accretion halted by the magnetic-barrier.)

The importance of accumulating of the magnetic flux
has been explored and observed by others in various astrophysical
contexts (e.g., Bisnovatyi-Kogan \& Ruzmaikin 1974, 1976;
Narayan, Igumenshchev \& Abramowicz 2003; INA03).
In particular, INA03 carried out a three-dimensional (3D)
MHD simulation (their model B) to late model times.
They found that the magnetic flux accumulates, initially
near the BH and then farther out, and the field
becomes dynamically dominant. At late times, mass is able
to accrete only via narrow streams, in a highly nonaxisymmetric
manner (see also Narayan et al. 2003).

The main difference between PB03's and INA03's results is the extent
and duration of the magnetic dominance. In PB03, the magnetic dominance 
is a {\em transient} whereas in INA03 is a {\em persistent} state.
The reason for this difference is the treatment of the magnetic field:
for the initial conditions, PB03 used the split-monopole magnetic field
and any changes in the magnetic flux near the BH during the evolution
are due to the chaotic, small-scale fields generated in the disc.
The detailed analysis show that the disc
properties in PB03's simulations are determined by MRI.  
In particular, MRI is responsible for
the complex field structure and
for the disc toroidal field being one or even two orders of magnitude
higher than the poloidal field (see figs. 9 and 10 in PB03 
and fig. 3 in Proga et al. 2003).
On the other hand, in their model B, INA03 set up a poloidal
field configuration in the injected gas in such a way that the portion
of the material that accretes always carries in the same sign
of the vertical component of the magnetic field.
The simulations carried out by PB03 and INA03 differ also 
in the assumed geometry (axisymmetric versus fully 3D). 
INA03 and PB03 
do not explore all cases including the case where the external or initial 
field has zero net flux or the field with the poloidal component changing 
sign on length scales much smaller than the size of 
the mass reservoir \footnote{ 
In the case where the initial or external flux has zero-net flux, 
a large scale coherent field might in some circumstances be generated by MRI  
(e.g., Livio, Pringle, \& King  2003). If so the central magnetic flux
could vary with time but still be dynamical signifacant for
some periods of time.}. Additionally, these simulations
also do not give definitive answers to the problems for which they were
designed.  Nevertheless, they give interesting insights 
into the general problem of MHD accretion flows. In particular, 
they suggest that magnetic
fields can provide an important parameter determining the time scale for
the accretion; i.e., it can be significantly longer than 
the local dynamical time scale.
This can have important implications for the observed X-flares in GRBs,
as we argue here, and for X-ray spectral states for BH binaries
as discussed by Spruit \& Uzdensky (2005, SU05 hereafter).
In fact, the work by SU05 describes very well the general physics and theory
of magnetic flux accumulated by an accretion flow.
Therefore we now turn our attention to some theoretical aspects of the problem
as presented by SU05.

\subsection{Theory of the magnetic barrier and accretion flow}

SU05 considered a new mechanism of efficient inward transport of
the large-scale magnetic field through a turbulent accretion disc.
The key element of the mechanism is concentration of the external field
into patches of field comparable in strength to  the MRI turbulence
in the disc. They focused on how to increase the magnetic flux at the center
in the context of BH binaries. In particular, they argue that the capture of
external magnetic flux by accretion disc and its subsequent compression
in the inner regions of the disc may explain both changes in the radiation
spectrum and jet activity in those objects.
However, their model and physical arguments are generic
and applicable to our problem.

One can expect that as the strength of the magnetic field
increases at the center, the field may eventually suppress MRI turbulence
and reduce the mass accretion rate and the power in the outflow.
This should be the case especially for GRBs because the mass inflow
rate at the late time is most likely much lower than at the early time.
The disc may become a Magnetically-Dominated Accretion Flow (MDAF)
as proposed by Meier (2005) or the fields in the polar funnel can expand
toward the equator and reconnect as in PB03's simulations.
In the latter, the torus is pushed outward by the magnetic field.
At this time, the gas starts to pile up outside the barrier;
eventually it can become unstable to interchange instabilities
at the barrier outer edge as suggested by SU05 or  the gas in the torus
can squash the magnetic field (compare
Fig.~5 and 6 in PB05 or the middle and right panel in Fig.1 here).

When interchange instabilities operate, magnetic flux from the bundle mixes
outward into the disc while the disc material enters the barrier.
In the accretion disk context, interchange instabilities 
have been studied by a few authors
(e.g., Spruit et al. 1995; Lubow \& Spruit 1995;
Stehle 1996; Stehle \& Spruit 2001; Li \& Narayan 2004). 
These studies showed that the onset of small-scale modes typical 
of interchanges (as in Rayleigh-Taylor instabilities) takes place only 
at rather large field strengths, due to a stabilizing effect of the Keplerian 
shear. The interchange instability operates at moderate field strengths,
but only at low shear rates (less than Keplerian). However for most of 
the time, we expect high shear rates in a torus because a low shear torus 
quickly becomes Keplerian due to MRI (e.g., PB03 and Proga et al. 2003).
We note that SU05 interpreted INA03 accretion through the barrier,
in the form of blobs and streams as a product of interchange instabilities.

SU05 also suggested that the field strength at which these instabilities 
become effective is most usefully expressed in terms of the degree of support
against gravity provided by the magnetic stress $B_R B_Z$.
According to SU05, the instabilities become effective when the radial magnetic
force, $F_m\sim 2 B_R B_Z/4\pi$, is of the order of a few percent
of the gravitational force, $F_g=GM\Sigma/R^2$, where $M$ is the central mass,
$R$ is the radius, and $\Sigma$ is the surface density.
For $B_R \approx B_Z$, there is a range in field strengths between
the value at which MRI turbulence is suppressed and the value where dynamical
instability of the barrier itself sets in, where no known instability operates
(Stehle \& Spruit 2001). In this range, the disk material
cannot mix or penetrate the magnetic field accumulated at the center
(e.g., the middle panel of Fig. 1).
Instead, mass builds up outside a region with such field
strengths until the magnetic field at the center is compressed enough
for instability to set in.

Thus, both numerical work and theoretical models of
magnetized accretion flows show that the inner most
part of the flow and accretor can respond {\em actively} to
changes of the accretion flow at larger radii.
In particular, the inner most accretion flow can be halted
for a very long time as shown by INA03 or it can be
repeatedly halted and reactivated as shown in PB03.

\subsection{Analytic estimates}

We finish this section with order-of-magnitude estimates of a few key features
of our X-ray flare model. We start by estimating the strength and flux of
magnetic field required to support the gas.
The gas of the surface density, $\Sigma$ can be supported against
gravity by the magnetic tension if $F_g\sim F_m$.
The surface density can be estimated from
$\Sigma=\MDOT /2\pi R \epsilon v_{ff}~{\rm g~cm^{-2}}$,
where  $\epsilon v_{ff}$ is the flow radial velocity assumed to be
a fraction $\epsilon$ of the free fall velocity, $v_{ff}$.
Assuming $B_r \approx B_z=B$, the force balance yields the field strength
$B \sim 2\times 10^{16}~\epsilon_{-3}^{-1/2} r^{-5/4} \MDOT_1^{1/2} M_3^{-1}
$~G, where $\epsilon_{-3} \equiv 10^{3} \epsilon$,
$r \equiv R/R_S=R/(2GM_{BH}/c^2)$,
$\MDOT_1=\MDOT/1~\MSUN~{\rm s^{-1}}$, and
$M_3=M/3 \MSUN$. We estimate the magnetic flux as
$\Phi\sim\pi r^2 R_S^2 B(r)= 5\times10^{28}~\epsilon_{-3}^{-1/2}r^{3/4}\MDOT_
1^{1/2} M_3~{\rm cm^2~G}$ from which we obtain an estimate
to the magnetospheric radius
$r_m \approx 60~\epsilon_{-3}^{2/3} \MDOT_1^{-2/3} M_3^{-4/3} \Phi_{30}^{4/3}$,
where $\Phi_{30}\equiv \Phi/(10^{30}~{\rm cm^2 G})$.
Substituting the expression for $B$ into the expression for the surface
density, one finds that a given magnetic flux can support
the gas with the surface density of
$\Sigma_B=5\times10^{19}~\Phi_{30}^2 M_3^{-3}r^{-2}~{\rm g~cm^{-2}}$.

To stop accretion with the hyper rate of $1\MSUN~{\rm s^{-1}}$ onto
a 3$\MSUN$ black hole at r=3 (i.e., for $r_m$ to be 3),
the magnetic flux of order $\Phi_{30} \sim 0.11$ is required.
We now assume that such a magnetic flux is accumulated during hyperaccretion
and that it does not change with time. Under these assumptions, $r_m=300$ for
the mass supply rate of $10^{-3}~\MSUNYR$ representative of
the late time evolution . 
This relatively large radius demonstrates one of
our key points that the innermost part an accreting system
can actively respond, via magnetic fields,
to changes in the inflow at large radii.

To estimate the conditions needed to restart accretion,
the accretion energetics and related time scales, we ask what is the mass of
a disc with $\Sigma$ high enough to reduce $r_m$ from 300 to 3 or so.
To answer this question, we adopt Popham et al.' (1999) model
of neutrino-dominated discs. Popham et al. assumed that
neutrino cooling produces a thin disc (Shakura \& Sunyaev 1973)
for accretion rates require to power GRBs.
Using  the disc solution for the
density and height (eqs. 5.4 and 5.5 in Popham et al. 1999),
we can express the disc surface density as
$\Sigma_\alpha=1.8\times10^{19}~\alpha^{-1.2} M_3^{-0.8}\MDOT_1r^{-1.25}$~g,
where $\alpha$ is the dimensionless parameter
scaling the stress tensor and the gas pressure (Shakura \& Sunyaev 1973).
Equating $\Sigma_B$ with $\Sigma_\alpha$,
one can estimate the mass accretion rate of an $\alpha$ disc
and compute $M_D$ by integrating $\Sigma_\alpha$ over radius.
For $\Phi_{30}=0.11$ and $\alpha=10^{-2}$
the accretion rate through the $\alpha$ disc is  $0.03~\MSUN~s^{-1}$
and $M_D$  for $r$ between 3 and  300  is 0.32~$\MSUN$. This
mass accretion rate is more than one order of magnitude  lower
than the rate of  $\sim 1~\MSUN~s^{-1}$ typical for 
the early time evolution. Thus, our estimates
are consistent with the fact that the X-ray flare luminosity
is at least one or two orders of magnitude lower the prompt 
gamma-ray emission (see section 3).
If this disc mass is a result of slow mass accumulation during
the late evolutionary stage, then it will take about 400 s to
rebuild the disc for the mass supply rate of $10^{-3}~\MSUN~s^{-1}$
and 12 s to accrete all this mass at the disc accretion rate of
$0.03~\MSUN~s^{-1}$. The latter is a lower estimate for
the flare duration because, for simplicity, we assumed a relatively high,
{\em constant} disc accretion rate. It is very likely that the rate
changes with time as the shape of the light curve
during the flares indicates.
In our model, the mass supply rate controls
the epochs when the flares happen:
the disc is rebuilt on the time scale which increases with time
because the mass supply slowdowns. Additionally, the flare duration
is coupled to the epoch through  the mass of the rebuilt disc.
Thus our model is capable of accounting for the observed
duration - time scale correlation.

\section{Discussion and conclusions}

The detailed analysis of the X-ray flares revealed that
they generally have lower luminosities (by at least one or two orders
of magnitude) than the prompt gamma-ray emission. Additionally,
the total energy of the flare is also typically smaller than that of
the prompt emission, although in some cases both could be comparable
(e.g. for GRB 050502B, Falcone et al. 2006).
Moreover multiple flares are observed in some GRBs and
the durations of these flares seem to be positively correlated with
the epochs when the flares happen, i.e. the later the epoch,
the longer the duration (O'Brien et al. 2005; Falcone et al. 2006;
Barthelmy et al. 2005).
The flare analysis also showed that the later the epoch
the lower the flare luminosity.
The above qualitative properties of the flares provide important constraints
on models of them.

Perna et al.'s (2006) disc fragmentation model promises to
account for the duration - time scale correlation and the duration - peak
luminosity anticorrelation. However, the physical process or processes
causing fragmentation are uncertain. It is also uncertain that
the conditions for the disc fragmentation are met in GRB progenitors.
This seems to be the case especially for the collapsar
model as a relatively high rotation of the progenitor
is required. We also note that magnetic fields can suppress or even prevent
disc fragmentation (e.g., Banerjee \& Pudritz 2006).

Here, we propose that the X-ray flares in GRBs are consequences
of the fact that during the late time evolution of a hyperaccretion
system the mass supply rate should decrease with time while
the magnetic flux accumulating around a BH should increase.
In particular, we point out that the flux accumulated during
the main GRB event can change the dynamics of the inner accretion flow.
We argue that the accumulated flux is capable of
halting intermittently the accretion flow. In our model,
the episode or episodes when the accretion resumes correspond to
X-ray flares.
A comparison of our analytic estimates from Section 2.3
with the observed X-ray flare characteristics, shows that our model is
not only physically based but also can both qualitatively and quantitatively
account for some aspects of the flares -- such as the peak times.
In general, our model fits under the general label of
the magnetic jet model for GRBs as we appeal to the magnetic effects to play
the key role not only during the main event but also during the late evolution.
The importance of the magnetic effects for the X-ray flares can be argued based
on energy budget of the accretion model (Fan et al. 2005).

The X-ray flares discovered in GRBs are relatively new and unexpected
phenomena. They give a strong incentive to apply the existing models
of hyperaccretion systems to circumstances where the mass supply is reduced.
Studies of this kind should reveal whether  one  needs
to introduce additional physics in order
to explain the flares. If so one should
explore the effects of this on the early evolution of GRBs
and check whether they are consistent with GRBs observations.
Our X-ray flare model has the advantage that it is essentially the same
as the MHD collapsar model for GRBs, with
only one justifiable change in a
key physical property of the collapsar model:
a decrease of the mass supply rate with time.

\section*{Acknowledgments}
We thank D. Meier, D. Uzdensky, and a referee for useful comments.
This work is supported by NASA under grants
NNG05GB68G (DP) and  NNG05GB67G (BZ).

\bsp

\begin{figure*}
\begin{picture}(180,180)
\put(-300,200){\includegraphics{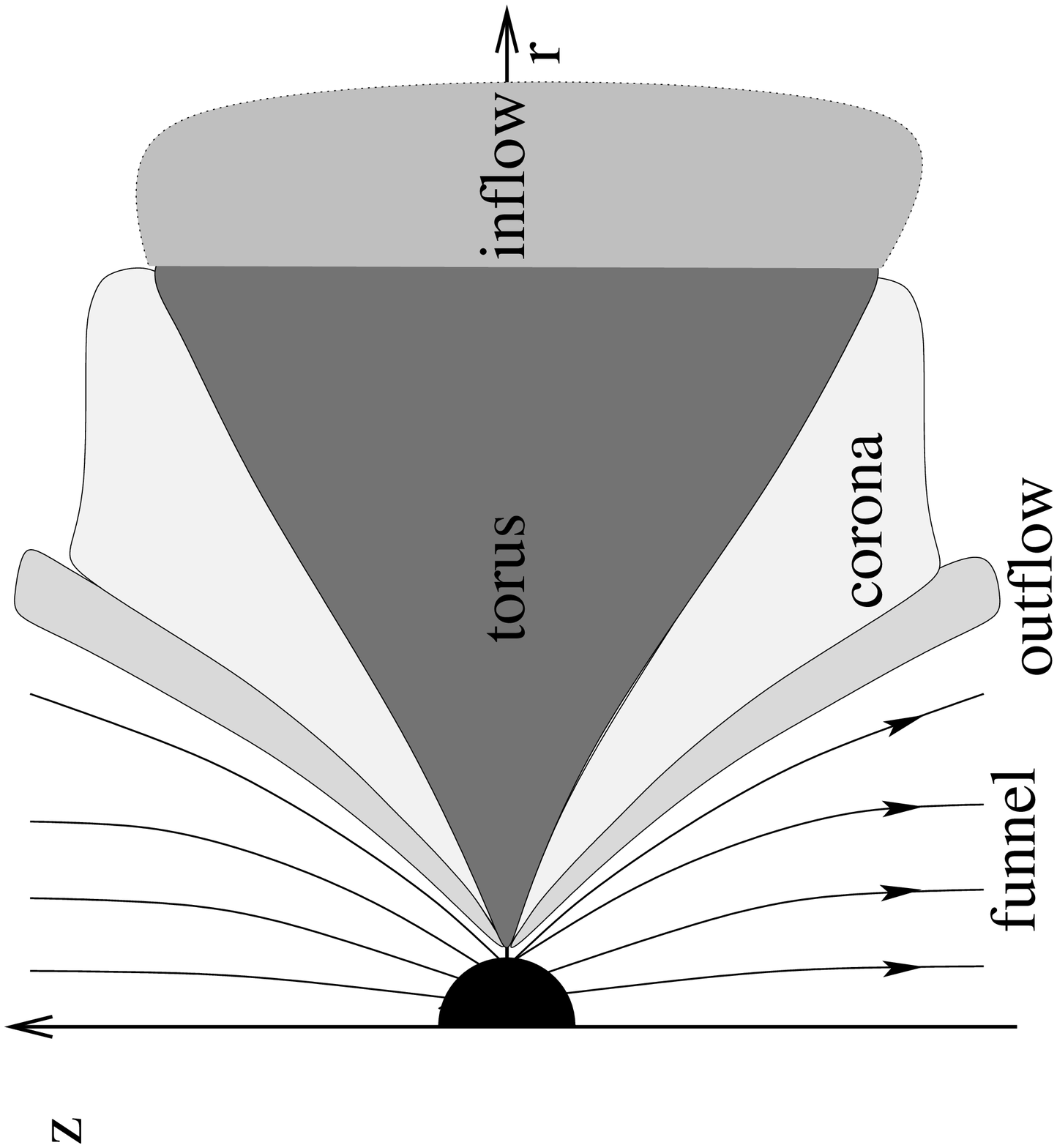}}
\put(-140,200){\includegraphics{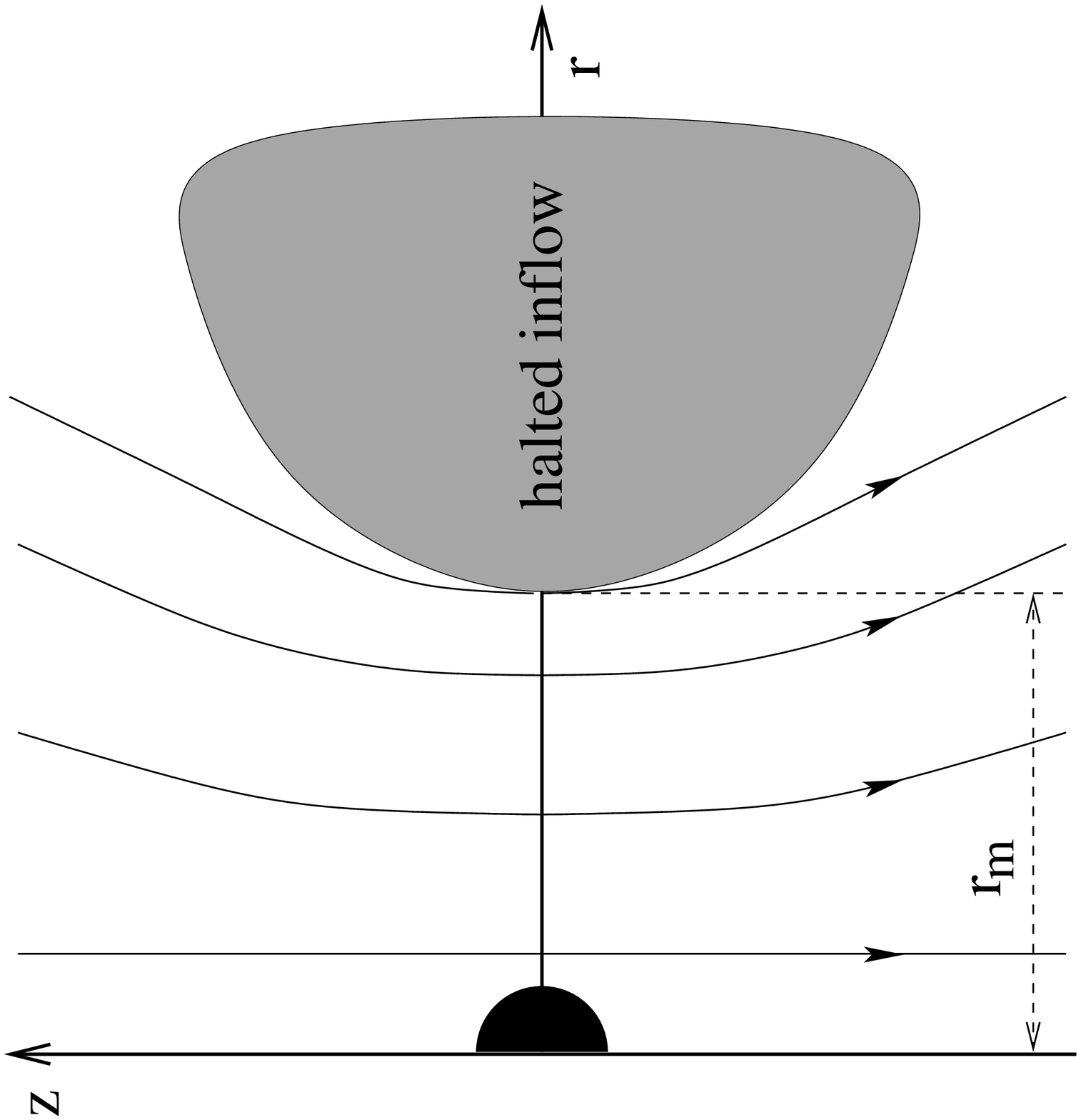}}
\put(20,200){\includegraphics{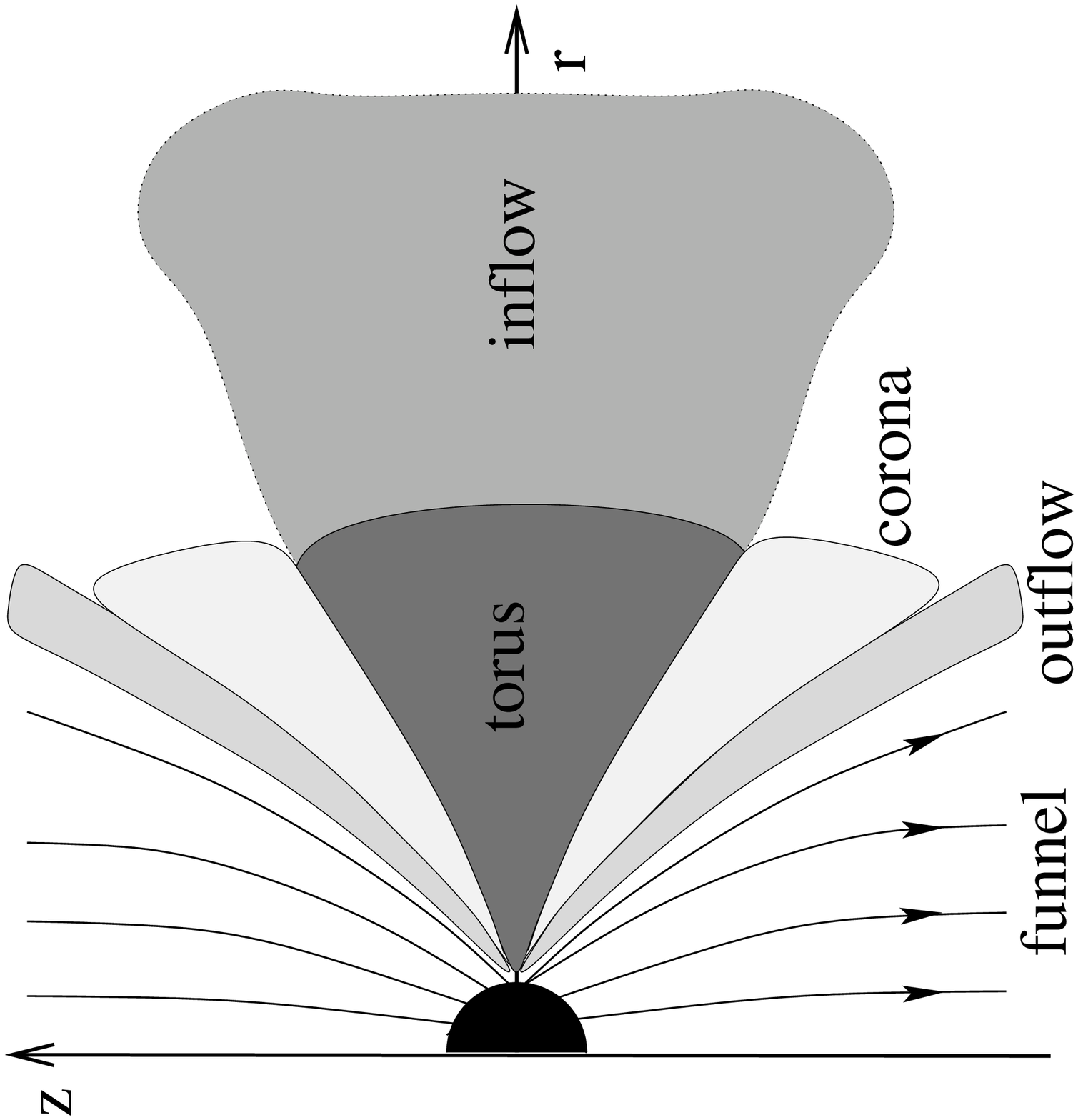}}
\end{picture}
\caption{ From left to right: General structural features of
the inner MHD flow  at three different accretion stages:
1) the inner flow during the hyperaccretion when
a relativistic jet forms and a strong poloidal magnetic
field is being accumulated at the center. The hyperaccretion
can not be sustained because the mass supply rate from the outer inflow
drops with time (hence different shades used for the torus and inflow).
2) The flow when
the hyperaccretion ended and the inflow rate is relatively low.
During this stage the magnetic field that accumulated earlier, can support
the gas against gravity. Consequently the inflow almost stops
at the distance comparable to the magnetospheric radius, $r_m$.
This stage ends when the surface density of the flow is too high
for the magnetic field to support the gas.
3) The inner flow when the magnetosphere is
squashed by the gas accumulated in the front of the inflow.
The accretion torus is rebuilt and a relativistic jet is reproduced.
The accretion rate at this stage is lower than
the hyperaccretion rate but higher than the inflow rate.
We expect that during the late accretion the inner flow
switches between the second and third stage and the third
stage corresponds to the time when X-ray flares are produced.
This cartoon illustrates the situation when the central magnetic flux
is conserved (i.e., the solid lines with arrows correspond to the magnetic
field lines at the center; for the clarity of the cartoon, the magnetic
field lines of other flow components are not drawn.)}
\end{figure*}


\end{document}